\documentclass[a4paper,12pt]{article}
\begin{document}
\rm
\begin{center} 
{\large {\bf Partial-wave Coulomb transition matrices\\
for attractive interaction by Fock's method}} \\[.2in]
{\bf  V. F. Kharchenko} \\[.05in]
{\small {\it Bogolyubov Institute for Theoretical Physics, \\ 
National Academy of Sciences of Ukraine, UA - 03143, Kyiv, Ukraine}} \\ 
{\footnotesize {\it vkharchenko@bitp.kiev.ua}} \\ [.1in]
\end{center}
{\small Leaning upon the Fock method of the stereographic projection of the 
three-dimensional momentum space onto the four-dimensional unit sphere the possibility 
of the analytical solving of the Lippmann-Schwinger integral equation for the partial   
wave two-body Coulomb transition matrix at the ground bound state energy 
has been studied. In this case new expressions for the partial $p$-, $d$- and $f$-wave 
two-body Coulomb transition matrices have been obtained in the simple 
analytical form. The developed approach can also be extended to determine analytically
the partial wave Coulomb transition matrices at the energies of excited bound 
states.} \\ [.05in]
{\small {\it Keywords:} Partial wave Coulomb transition matrix; Lippmann-Schwinger equation;
Fock method; Analytical solution \\ [.05in]
}  \\ [.2in]
\noindent {\bf 1. Introduction} \\ 

It is well known that the transition matrix of two interacting particles (with 
its elements both on and off the energy shell) encloses all the information on 
the system. In view of this, the transition matrix is an involved function of the 
variables of the initial and final momenta and the energy. In particular, the 
possible bound states of the system are responsible for formation in the off-shell 
$t$-matrix of the corresponding poles in the energy with the residues related with 
the wave functions of the system in these states. Similarly, providing the property 
of the unitarity of the scattering matrix, the states of the continuum spectrum 
at positive energies manifest themselves in existence in the transition matrix 
of a well-known singularity with a branch point and a cut along the positive 
energy axis. 

A knowledge of the two-body off-shell Coulomb transition matrix is of 
great interest being necessary for description of the few-body 
atomic and nuclear systems. In particular, the Coulomb transition matrices are 
contained in the kernels of the Faddeev integral equations [1,2] for three 
particles, among which two or all particles are charged.

For charged particles, several representations of the two-body Coulomb transition 
matrix $t^C(E)$ (or the related Coulomb Green function $g^C(E)$) are known [3-12]. 
A considerable body of the earlier information on the Coulomb $t$-matrix is 
contained in the overview [13].

Of paricular importance are the studies of the Coulomb transition matrix with 
the use of the symmetries of the Coulomb system both in the usual 
three-dimensional and Fock's four-dimensional [14] Euclidean spaces 
(Refs. [5] and [8]). In this way the expressions for the Coulomb transition 
matrix with the explicit removal of the sigularities in the transfer-momentum 
variable $\mid {\bf k}-{\bf k^{\prime}}\mid$ and the energy $E$ have been derived 
in the papers [10] (for the negative energy $E<0$) and [11] (for the 
zeroth and positive energies $E\geq 0$).

This paper is devoted to the derivation of the partial wave two-body 
Coulomb transition matrices in the special case that the energy $E$ is equal to 
the qround bound state energy of the complex. In Section 2 we start from the 
expression for three-dimensional Coulomb transition matrix at the negative energy 
obtained by us (in Ref.[10]). In Section 3 we formulate the method 
for solving the one-dimensional integral Lippmann-Schwinger equation for the partial 
wave Coulomb $t$-matrix at the negative energy $E<0$. In Section 4 the general formula for 
the partial wave Coulomb $t$-matrix at the energy of the ground bound state $E=-b_1$ 
is derived. The analytical expressions for the $p$-, $d$- and $f$-wave components 
of the Coulomb $t$-matrix are given in Subsections 4.1 -- 4.3. Section 5 is 
devoted to the concluding remarks.
\\ [.2in]
\noindent {\bf 2. Three-dimensional Coulomb transitions matrix at negative energy}\\

The two-particle Coulomb Green operator $g^C(E)\equiv (E-h_0-v^C)^{-1}$ and the transition 
operator $t^C(E)$,which describe the system of two charged particles 1 and 2, are related 
among themselves by the relation
\begin{equation}
g^C(E)= g_0(E) + g_0(E) t^C(E) g_0(E) \;.
\end{equation}
Here $h_0$ denotes the operator of the kinetic energy of the relative motion of the 
particles, $g_0(E)\equiv (E-h_0)^{-1}$ is the free Green operator, $v^C(r)=q_1 q_2 / r$ 
is the potential of the Coulomb interaction between the particles ($q_i$ denotes the 
charge of the particle $i$, $r$ is the distance between the particles). The quantity $E$ 
denotes the total energy of the relative motion between the particles 1 and 2.

In the momentum representation the formula (1) has the form 
\begin{equation}
<{\bf k}|g^C(E)|{\bf k}^{\prime}>=<{\bf k}|g_0(E)|{\bf k}^{\prime}>+\frac{1}{E-\frac{k^2}{2\mu}}
<{\bf k}|t^C(E)|{\bf k}^{\prime}>\frac{1}{E-\frac{k^{{\prime}2}}{2\mu}}\;,
\end{equation}
where ${\bf k}$ and ${\bf k}^{\prime}$ are the relative momentum variables that correspond to the 
relative radius vectors ${\bf r}$ and ${\bf r}^{\prime}$ in the coordinate space, 
$\mu = m_1 m_2/(m_1 + m_2)$ is the reduced mass of the particles 1 and 2 (${\bf r}={\bf r}_1-{\bf r}_2$, 
${\bf r}_i$ and $m_i$ are the radius vector and the mass of the particle $i$),
\begin{displaymath}
<{\bf k}|g_0(E)|{\bf k}^{\prime}>=\frac{(2\pi)^3 \delta ({\bf k}-{\bf k}^{\prime})}{E-\frac{k^2}{2\mu}}\;.
\end{displaymath}
Three-dimensional Coulomb transition matrix $<{\bf k}|t^C(E)|{\bf k}^{\prime}>$ satisfies to the 
Lippmann-Schwinger integral equation
\begin{equation}
<{\bf k}|t^C(E)|{\bf k}^{\prime}>=\langle {\bf k} \mid v^C \mid {\bf k}^{\prime}\rangle + 
\int \frac{d{\bf k}^{\prime\prime}}{(2\pi)^3}\langle {\bf k} \mid v^C \mid {\bf k}^{\prime\prime}\rangle
\frac{1}{E-\frac{k^{\prime\prime 2}}{2\mu}} <{\bf k}^{\prime\prime}|t^C(E)|{\bf k}^{\prime}>\;\;,   
\end{equation}  
where the matrix of the potential of the Coulomb interaction between the particles 1 and 2 
is equal to
\begin{equation}
\langle {\bf k} \mid v^C \mid {\bf k}^{\prime}\rangle = 
\frac{4\pi q_1 q_2}{\mid {\bf k}-{\bf k}^{\prime}\mid ^2}\;. 
\end{equation}
 
At the negative energy
\begin{equation}
E = - \frac{\hbar^2 \kappa^2}{2\mu}<0
\end{equation} 
the solution of the equation (3) has the form [10]
\begin{displaymath}
 <{\bf k}|t^C(E)|{\bf k}^{\prime}>=\frac{4\pi q_1 q_2 \kappa^2}{(k^2+\kappa^2)
(k^{\prime 2}+\kappa^2)}\left[ \frac{1}{\sin^2\frac{\omega}{2}} 
-2\pi\gamma \frac{\cos\gamma\omega}{\sin\omega}
-2\gamma\frac{\sin 2\gamma \omega}{\sin \omega}\ln (\sin\frac{\omega}{2}) \right.
\end{displaymath}
\begin{equation}
 + 4\pi\gamma\; c(\gamma)\; \cot \gamma\pi \frac{\sin\gamma\omega}{\sin\omega} 
+ 2 \gamma \frac{\cos \gamma \omega}{\sin \omega}\int_{0}^{\omega} d\varphi \;
\sin \gamma\varphi\; cot\frac{\varphi}{2} 
\end{equation} 
\begin{displaymath}
\left.  + 4\gamma^2 \frac{\sin\gamma\omega}{\sin\omega} 
\int_{\omega}^{\pi} d\varphi \; \sin \gamma\varphi \;\ln(\sin \frac{\varphi}{2})\right],
\end{displaymath}
where $\gamma$ is the dimensionless quantity that denotes the Coulomb parameter
\begin{equation}
\gamma = \frac{\mu q_1 q_2}{\hbar^2 \kappa}\;,
\end{equation} 
$\hbar$ is the reduced Planck's constant. The variable quantity $\omega$ in Eq.(6) 
refers to the angle between two 4-vectors $e\equiv ({\bf e}, e_0)$ and $e^{\prime}\equiv 
({\bf e{\prime}}, e_0^{\prime})$ in a four-dimensional Euclidean space 
introduced by Fock [14],
\begin{equation}
{\bf e}=\frac{2\kappa {\bf k}}{\kappa^2 + k^2}\;,\quad 
e_0 = \frac{\kappa^2 - k^2}{\kappa^2 + k^2}\;; \qquad 
{\bf e}^{\prime}=\frac{2\kappa {\bf k}^{\prime}}{\kappa^2 + k^{\prime 2}}\;,\quad 
e_0^{\prime} = \frac{\kappa^2 - k^{\prime 2}}{\kappa^2 + k^{\prime 2}}\;; 
\end{equation} 
\begin{equation}
\cos \omega = e\cdot e^{\prime} = {\bf e}\cdot {\bf e}^{\prime} + e_0\cdot 
 e_{0}^{\prime}\;.
\end{equation} 
In this case, the vectors ${\bf k}$ and ${\bf k}^\prime$ lie in the hyperplane, which is 
the stereographic projection of the unit sphere in the four-dimensional space, and the 
variable $\omega$ is defined by the expression
\begin{equation}
\sin^2\frac{\omega}{2} = \frac{{\kappa^2}\mid {\bf k} - {\bf k}^\prime \mid ^2}
{(k^2 + \kappa^2)(k^{\prime 2} + \kappa^2)}\;\; , \;\; 0\leq \omega \leq \pi\;\;.
\end{equation} 
The function $c(\gamma)$ in Eq.(6) has the form
\begin{equation}
c(\gamma)= \frac{1}{2} \left( 1 - \frac{1}{\pi}\int_{0}^{\pi} d\varphi \;\sin \gamma\varphi 
\;\cot \frac{\varphi}{2} \right)
\end{equation}
or using the digamma functions
\begin{equation}
c(\gamma)= \theta(-\gamma)+\frac{\sin \gamma\pi}{2\pi} \left[ \psi \left( \frac{\mid \gamma\mid +1}
{2} \right)- \psi \left( \frac{\mid \gamma\mid }{2} \right) - \frac{1}{\mid \gamma\mid } \right]\;,
\end{equation}
where $\psi(x)\equiv d/dx \ln\Gamma(x)$ and $\Gamma(x)$ are the digamma and gamma functions [15], 
and $\theta(x)$ is the usual step function
\begin{equation}
\theta(x)=\left\{ \begin{array}{cc}
                 1 & \mbox{for $x>0$}\;, \\
                 0 & \mbox{for $x<0$}\;.
                 \end{array}
          \right.
\end{equation}

The first three terms of the expression in brackets (6) contain singularities in the 
transfer momentum:
\begin{displaymath}
\mid {\bf k}-{\bf k}^{\prime}\mid ^{-2}\;, \quad \mid {\bf k}-{\bf k}^{\prime}\mid ^{-1} \;
\mbox{ and }\; \ln \left\{\kappa \mid {\bf k}-{\bf k}^{\prime}\mid 
/(k^2 + \kappa^2)^{1/2}(k^{\prime 2} + \kappa^2)^{1/2}\right\}
\end{displaymath}
respectively in the first, second and third terms. The other three terms in (6) are 
smooth functions of $\mid {\bf k} - {\bf k}^{\prime}\mid $. 

The singularities in the energy are contained in the fourth term of the expression (6). They arise 
only in the case of the attractive Coulomb interaction potential (with unlike electric charges, 
$\gamma<0$) when the Coulomb parameter $\gamma$ takes on the negative integer values ($\gamma 
= -n,\; n=1,2,3,...$) and correspond to the bound states. At this points $cot \gamma \pi$ has the 
pole singularities and the function $c(\gamma)$ is distinct from zero: $c(-n)=1$. In the case of 
the repulsive Coulomb potential ($\gamma > 0$) for the positive integer value of $\gamma$ the 
function vanishes, $c(n)=0$, and the fourth term in (6) is finite:
\begin{equation}
\lim_{\gamma \rightarrow n}\pi c(\gamma) \cot \gamma \pi =  
\frac{(-1)^n}{2n} - \sum_{m=1}^{n} \frac{(-1)^m}{m} - \ln 2\;.
\end{equation}

The fifth and sixth terms in the expression (6) are the smooth functions of the energy 
(or of the Coulomb parameter $\gamma$).
\\ [.2in]
\noindent {\bf 3. Partial wave decomposition of the Coulomb transitions matrix}\\

For the Coulomb potential (4), which is local and spherically symmetric, the expansions of the 
matrix elements $\langle {\bf k} \mid v^C \mid {\bf k}^{\prime}\rangle$ and 
$\langle {\bf k} \mid t^C(E) \mid {\bf k}^{\prime}\rangle$ in partial waves are 
\begin{displaymath}
\langle {\bf k} \mid v^C \mid {\bf k}^{\prime}\rangle = \sum_{l=0}^{\infty} (2l+1) 
v_l^C(k,k^{\prime}) P_l(\hat{{\bf k}}\cdot \hat{{\bf k}}^{\prime})\;,
\end{displaymath}
\begin{equation}
\langle {\bf k} \mid t^C(E) \mid {\bf k}^{\prime}\rangle = \sum_{l=0}^{\infty} (2l+1) 
t_l^C(k,k^{\prime};E) P_l(\hat{{\bf k}}\cdot \hat{{\bf k}}^{\prime})\;.
\end{equation}
Here $P_l(x)$ is the Legendre polinomial, the quantity with the hat $\hat{\bf k}$ 
denotes the unit vector along ${\bf k}$ and depends on the polar angles of ${\bf k}$, 
$\hat{\bf k}\cdot \hat{{\bf k}}^{\prime}=\cos \theta$ .

The elements of the partial Coulomb potential and transition matrices are
\begin{displaymath}
v_l^C(k,k^{\prime}) = \frac{1}{2} \int_{0}^{\pi} d\theta \;\sin \theta \;P_l(\cos \theta)
\langle {\bf k} \mid v^C \mid {\bf k}^{\prime}\rangle \;,
\end{displaymath}
\begin{equation}
t_l^C(k,k^{\prime};E) = \frac{1}{2} \int_{0}^{\pi} d\theta \;\sin \theta \;P_l(\cos \theta)\;
\langle {\bf k} \mid t^C (E) \mid {\bf k}^{\prime}\rangle \;,
\end{equation}

The partial component of the Coulomb potential (4) is 
\begin{equation}
v^C_l(k,k^{\prime}) = \frac{2\pi q_1 q_2}{k k^{\prime}} Q_l 
\left( \frac{k^2+{k^{\prime}}^2}{2 k k^{\prime}} \right)\;,
\end{equation}
where the function $Q_l(x)$ is the Legendre function of the second kind [15] 
\begin{equation}
Q_l(x) = \frac{1}{2} P_l(x) \ln \left( \frac{x+1}{x-1}\right)- W_{l-1}(x)\;,
\end{equation}
\begin{displaymath}
W_{-1}(x)=0\;, \qquad  W_{l-1}(x)=\sum_{k=1}^{l} \frac{1}{k} P_{l-k}(x) P_{k-1}(x)\;.  
\end{displaymath}

The partial wave component of the Coulomb transition matrix $t_l^C(k,k^{\prime};E)$ 
satisfies the one-dimensional integral equation that follows from (3)
\begin{equation}
t^C_l(k,k^{\prime};E) = v^C_l(k,k^{\prime})+
\int_{0}^{\infty} \frac{dk^{\prime\prime} {k^{\prime\prime}}^2}{2\pi^2}
v^C_l(k,k^{\prime\prime}) \frac{1}{E-\frac{{k^{\prime\prime}}^2}{2\mu}}
t^C_l(k^{\prime\prime},k^{\prime};E)\;\;, \\[3mm]  
\end{equation}  

We find the solution of the equation (19) in the case of the negative energy $E<0$ by 
substituting the expression for the three-dimensional Coulomb transition matrix (6) 
into the formula for the partial transition matrix (16). Going from the integration 
variable $\theta$ (the angle between the three-dimensional vectors ${\bf k}$ and 
${\bf k}^{\prime}$) in (16) to the variable $\omega$ using the formulas following from 
the expression (10),
\begin{equation}
\cos {\theta} = \frac{\xi}{\eta} - \frac{1}{\eta} {\sin}^2{\frac{\omega}{2}} = 
\frac{2\xi - 1 + \cos {\omega}}{2\eta}
\qquad \qquad \sin {\theta} \; d\theta = \frac{1}{2\eta} \sin {\omega} \; d\omega \;,   
\end{equation}  
where
\begin{equation}
\xi = \frac{\kappa^2 (k^2 + {k^{\prime}}^2)}{(k^2 + \kappa^2)({k^{\prime}}^2 + \kappa^2)}\;,\quad 
\eta = \frac{2 \kappa^2 k k^{\prime}}{(k^2 + \kappa^2)({k^{\prime}}^2 + \kappa^2)}\;, 
\end{equation}  
we obtain 
\begin{equation}
t_l^C(k,k^{\prime};E) = \frac{1}{4\eta} \int_{\omega_0}^{\omega_{\pi}} d\omega\; 
\sin \omega \; P_l \left( \frac{2\xi-1+ 
\cos \omega}{2\eta} \right)
\langle {\bf k} \mid t^C (E) \mid {\bf k}^{\prime}\rangle  \;.
\end{equation}
The integration limits in (22) are determined by the expressions
\begin{equation}
\omega_0 = 2 \arcsin \sqrt{\xi - \eta}\;, \qquad \; \omega_{\pi} = 2 \arcsin \sqrt{\xi + \eta}\;,
\end{equation} 
in this case 
\begin{displaymath}
\cos \omega_0 = 1 - 2 \xi + 2 \eta\;, \qquad \; \cos \omega_{\pi} = 1 - 2 \xi - 2 \eta\;,
\end{displaymath}
\begin{equation}
\sin \omega_0 = 2 \sqrt{\xi - \eta} \sqrt{1 - \xi + \eta}\;, \qquad \; 
\sin \omega_{\pi} = 2 \sqrt{\xi + \eta} \sqrt{1 - \xi - \eta}\;.
\end{equation} 
Using the expression for the three-dimensional transition matrix (6) we write the formula 
(22) for the partial wave Coulomb transition matrix $t^C_l(k,k^{\prime};E)$ at $E<0$ in 
the form
\begin{displaymath}
t^C_l(k,k^{\prime};E)= \frac{\pi q_1 q_2}{k k^{\prime}} \int_{\omega_0}^{\omega_{\pi}}
d\omega \;P_l \left( \frac{2\xi-1+ \cos \omega}{2\eta}\right) 
\left\{ \cot {\frac{\omega}{2}} \right.
\end{displaymath}
\begin{equation}
- \pi \gamma \; \cos \gamma \omega - \gamma \;\sin 2\gamma\omega \;\ln (\sin \frac{\omega}{2}) 
+ 2\pi \gamma\; c(\gamma)\; \cot \gamma\pi\; \sin \gamma\omega  
\end{equation} 
\begin{displaymath}
\left. + \gamma \;\cos \gamma \omega \int_{0}^{\omega} d\varphi \;
\sin \gamma\varphi \;\cot \frac{\varphi}{2} + 2\gamma^2 \sin {\gamma\omega} 
\int_{\omega}^{\pi} d\varphi \;\sin \gamma\varphi \; \ln \left( \sin {\frac{\varphi}{2}}\right) 
\right\} \; .
\end{displaymath}
Notice that in the expression (25) the quantity $\gamma$ by the definition (7) depends on 
$\kappa$ or, in accordance with (5), on the energy $E$. The quantities $\xi$ and $\eta$, 
like the limits $\omega_0$ and $\omega_{\pi}$, depend on all the variables $k$, $k^{\prime}$ 
and $E$ of the partial transition matrix $t^C_l(k,k^{\prime};E)$.
\\ [.2in]
\clearpage
\noindent {\bf 4. Partial-wave Coulomb transition matrix at the ground bound state energy}\\

The general expression for partial wave Coulomb transitions matrix $t^C_l(k,k^{\prime};E)$
(25) is simplified in special cases if the energy $E$ takes the value of the bound (ground 
or excited) state of the two-particle system,$E=E_n$, where $n$ is the principal quantum  
number (for each value $n$ the angular momentum of the bound state $l_0$ can take all integer 
values from 0 to $n-1$),
\begin{equation}
E_n = - \frac{\mu (q_1 q_2)^2}{2 \hbar^2 n^2}\;, \qquad  n=1, 2, \cdots \;,
\end{equation}
herewith according to (5) and (7),
\begin{equation}
\kappa = \kappa_n = \frac{\sqrt{-2\mu E_n}}{\hbar} = \frac{\mu \mid q_1 q_2 \mid }{\hbar^2 n}\;, 
\qquad \gamma = \gamma_n = - n\;.
\end{equation}

In this work we restrict the study of the Coulomb transition matrix (25) at the 
energy of the ground bound state of the two-particle complex of the two-particle ($n = 1$):
\begin{equation}
E=E_1=-b_1\;, \qquad b_1 =\frac{\hbar^2 \kappa_1^2}{2\mu}\;, \qquad \kappa=\kappa_1\equiv
\frac{\mu\mid q_1 q_2 \mid}{\hbar^2}\;, \qquad \gamma = \gamma_1 = -1\;.
\end{equation}

According to (25) and the relations of orthogonality and normalization for the 
Legendre polynomials, we find
\begin{equation}
\int_{\omega_0}^{\omega_{\pi}} d\omega\; \sin \omega\; P_l \left( \frac{2\xi-1+ \cos \omega}
{2\eta} \right) = 2\eta \int_{0}^{\pi} d\theta\; \sin \theta\; P_l (\cos \theta)=4\eta \delta_{l0} \;,
\end{equation}
and hence only s-wave partial component $t^C_0(k,k^{\prime};E)$ has the pole singularity 
in energy at the point $E=E_1=-b_1$ (that locates in the fourth term) of the expression in 
braces (25). All higher partial components $t^C_l(k,k^{\prime};E)$ with $l>0$ are non-singular 
at this point.

The expression for partial Coulomb transition matrices with $l>0$ at the energy of the ground 
bound state of the two-particle complex (28) we obtain using the expression (25) with 
$\gamma \rightarrow -1$. Herewith, we evaluate the indeterminate form of the type $\frac{0}{0}$, 
which comes into being after the integration with respect to $\omega$ with the fourth term of 
the expression in braces (25) at $\gamma \rightarrow -1$, according to the l'Hospital rule:
\begin{displaymath}
\lim_{\gamma \rightarrow -1} 
\frac{2 \pi \gamma \; c(\gamma) \int_{\omega_0}^{\omega_{\pi}} d\omega\; \sin \gamma \omega\; 
P_l \left( \frac{2\xi - 1+ \cos \omega}{2\eta} \right) }{\tan \gamma \pi}
\end{displaymath}
\begin{equation}
\rightarrow -2 \int_{\omega_0}^{\omega_{\pi}} d\omega\; \omega\; \cos \omega\; 
P_l \left( \frac{2\xi-1+ \cos \omega}{2\eta} \right)\;. 
\end{equation}

At the energy of the ground bound state ($\gamma = -1$) the integrals over $\varphi$ 
in the fiveth and sixth terms of the expression (25) are equal to
\begin{displaymath}
 \int_{0}^{\omega} d\varphi\; \sin \varphi \;\cot \frac{\varphi}{2} = \omega + \sin \omega\;,
\end{displaymath}
\begin{equation}
\int_{\omega}^{\pi} d\varphi\; \sin \varphi \;\ln \left( \sin \frac{ \varphi}{2} \right) = 
- \cos ^2 \frac{\omega}{2} - 2 \sin ^2 \frac{\omega}{2} \;\ln \left( 
\sin \frac{\omega}{2} \right)\;.
\end{equation}

In this case the expression in braces of the general expression for the partial
Coulomb transition matrix (25) acquires the form
\begin{equation}
\cot \frac{\omega}{2} + \pi \cos \omega - \omega \cos \omega - \sin \omega - 
2 \sin \omega \ln \left( \sin \frac{\omega}{2} \right)\;. 
\end{equation}
It should be noted that the fourth term in the expression (32) does not contribute 
to all partial components $t_l^C(k,k^{\prime};-b_1)$ with $l>0$ in view of the 
orthogonality relation (29).

As a result the expression (25) for the partial wave Coulomb transition matrix 
(with $l=1,2,3,... $) at the ground bound state energy ($\gamma = -1$) is 
simplified to the form
\begin{displaymath}
t^C_l(k,k^{\prime};-b_1)= \frac{\pi q_1 q_2}{k k^{\prime}} \int_{\omega_{01}}^{\omega_{\pi 1}}
d\omega\; P_l \left( \frac{2\xi_1 - 1 + \cos \omega}{2\eta_1}\right) 
\end{displaymath}
\begin{equation}
\cdot \left\{ \cot \frac{\omega}{2} + \pi \cos \omega - \omega \cos \omega -2 \sin \omega\;  
\ln \left( \sin \frac{\omega}{2} \right) \right\} \;,
\end{equation} 
where the quantities $\xi_1, \eta_1, \omega_{01}$ and $\omega_{\pi 1}$ are determined 
by the expressions for $\xi, \eta, \omega_{0}$ and $\omega_{\pi}$ according to the 
definitions (21) and (23) taken at the value $\kappa = \kappa_1$ that corresponds to 
the energy of the ground bound state (28),
\begin{displaymath}
\xi_1 = \frac{\kappa_1^2 (k^2 + {k^{\prime}}^2)}{(k^2 + \kappa_1^2)({k^{\prime}}^2 + 
\kappa_1^2)}\;,\quad \eta_1 = \frac{2 \kappa_1^2 k k^{\prime}}{(k^2 + 
\kappa_1^2)({k^{\prime}}^2 + \kappa_1^2)}\;, 
\end{displaymath}
\begin{equation}
\omega_{01} = 2 \arcsin \sqrt{\xi_1 - \eta_1}\;, \quad \; \omega_{\pi 1} = 
2 \arcsin \sqrt{\xi_1 + \eta_1}\;,
\end{equation} 

Notice that taking into account only the first term in the braces of the expression (33) 
corresponds to the Born approximation for the partial Coulomb transition matrix
\begin{equation}
t^{C,B}_l(k,k^{\prime};-b_1) = v^C_l(k,k^{\prime})
\end{equation} 
that is determined by the formulas (17) and (18).

Hereafter, using the spectroscopic notation, we designate the low partial waves 
corresponding to $l=0,1,2,3,...$ as $s,\;p,\;d,\;f,\;...$ waves.
\\ [.2in]
\noindent {\sl 4.1. Partial $p$-wave Coulomb transition matrix
 at the energy $E=-b_1$}\\

The formula for the $p$-wave Coulomb $t$-matrix 
at the ground bound state energy follows immediately after integration in the 
expression (33) with $l=1$. The separate four terms in the braces of the expression (33) 
make the following contributions to the $p$-wave $t$-matrix:
\begin{equation}
t^C_1(k,k^{\prime};-b_1) = \frac{\pi q_1 q_2}{k k^{\prime}} \left\{ {\cal P}_1 + {\cal P}_2 + 
{\cal P}_3 + {\cal P}_4 \right\}\;,
\end{equation} 
where
\begin{displaymath}
{\cal P}_1 = \frac{\xi_1}{\eta_1} \ln \frac{\xi_1 + \eta_1}{\xi_1 - \eta_1} - 2 \;,
\end{displaymath}
\begin{displaymath}
{\cal P}_2 = \frac{\pi}{4 \eta_1} \left[ \omega_{\pi 1} - \omega_{01} - \sin (\omega_{\pi 1} 
- \omega_{01}) \right] \;,
\end{displaymath}
\begin{equation}
{\cal P}_3 = 2 \xi_1 - 1 + \frac{1}{\eta_1} \left[ -\frac{1}{8} \left( \omega_{\pi 1}^2 - 
\omega_{01}^2 \right) \right.
\end{equation} 
\begin{displaymath}
\left. + \frac{1}{4} \left( \omega_{\pi 1} \sin \omega_{\pi 1} \cos \omega_{01} 
- \omega_{01} \cos \omega_{\pi 1} \sin \omega_{01} \right) \right] \;,
\end{displaymath}
\begin{displaymath}
{\cal P}_4 = 2 \xi_1 - \frac{1}{\eta_1} \left( \xi_1^2 - \eta_1^2 \right) 
\ln \frac{\xi_1 + \eta_1}{\xi_1 - \eta_1} \;.
\end{displaymath}

The first term in (36) is the Born approximation for the $p$-wave Coulomb $t$-matrix 
\begin{equation}
t^{C,B}_1(k,k^{\prime};-b_1) = v^C_1(k,k^{\prime}) = \frac{\pi q_1 q_2}{k k^{\prime}} {\cal P}_1\;.
\end{equation} 
The first and fourth addends in (36) contain the logarithmic functions. 

Adding up in (36) all four terms (37) we obtain the following expression for the 
$p$-wave Coulomb transition matrix at $E=-b_1$:
\begin{displaymath}
t^C_1(k,k^{\prime};-b_1) = \frac{\pi q_1 q_2}{k k^{\prime}} \left\{ 4 \xi_1 - 3 + 
\frac{1}{\eta_1} \left[ (\xi_1 - \xi_1^2 + \eta_1^2 ) \ln \left( 
\frac{\xi_1 + \eta_1}{\xi_1 - \eta_1} \right) \right. \right.
\end{displaymath}
\begin{equation}
+ \frac{1}{8} \left( \omega_{\pi 1} - \omega_{01} \right) \left( 2 \pi - \omega_{\pi 1} 
- \omega_{01} \right) - \frac{1}{4} \left( \pi - \omega_{\pi 1} \right) 
\sin \omega_{\pi 1} \cos \omega_{01}
\end{equation} 
\begin{displaymath}
\left. \left. + \frac{1}{4} \left( \pi - \omega_{01} \right) \cos \omega_{\pi 1} 
\sin \omega_{01} \right] \right\}\;.
\end{displaymath}
\\ [.2in]
\noindent {\sl 4.2. Partial $d$-wave Coulomb transition matrix 
 at the energy $E=-b_1$}\\

In a similar way, performing integration in the expression (33) with $l=2$, we obtain the formula 
for the $d$-wave Coulomb $t$-matrix at the ground bound state energy:
\begin{equation}
t^C_1(k,k^{\prime};-b_1) = \frac{\pi q_1 q_2}{k k^{\prime}} \left\{ {\cal D}_1 + {\cal D}_2 +  
{\cal D}_3 + {\cal D}_4 \right\}\;.
\end{equation} 
The separate terms in (40) correspond to the addends in the braces of the expression (33)
for the $d$-wave component of the $t$-matrix,
\begin{displaymath}
{\cal D}_1 = \left[ \frac{3}{2} \left( \frac{\xi_1}{\eta_1} \right) ^2 - \frac{1}{2} \right] 
\ln \left( \frac{\xi_1 + \eta_1}{\xi_1 - \eta_1} \right) - 3 \frac{\xi_1}{\eta_1} \;,
\end{displaymath}
\begin{displaymath}
{\cal D}_2 = \frac{1}{\eta_1^2} \left\{ \frac{3\pi}{8} (2\xi_1 - 1) (\omega_{\pi 1} - \omega_{01} ) 
+ \frac{\pi}{8} \left[ X_2 (\xi_1,\eta_1) \sin \omega_{\pi 1} - X_2 (\xi_1,- \eta_1) 
\sin \omega_{01} \right] \right\} \;,
\end{displaymath}
\begin{equation}
{\cal D}_3 = \frac{1}{\eta_1} \left( 2 \xi_1^2 - 2 \xi_1 -\frac{4}{3} \eta_1^2 + 
\frac{3}{2} \right) - \frac{1}{\eta_1^2} \left[ \frac{3}{16} (2\xi_1 - 1) 
(\omega_{\pi 1}^2 - \omega_{01}^2 ) \right.
\end{equation} 
\begin{displaymath}
+ \left. \frac{1}{8} X_2 (\xi_1,\eta_1) \omega_{\pi 1} \sin \omega_{\pi 1} - \frac{1}{8} 
X_2 (\xi_1,- \eta_1) \omega_{01} \sin \omega_{01} \right]  \;,
\end{displaymath}
\begin{displaymath}
{\cal D}_4 = \frac{1}{\eta_1} \left( 2 \xi_1^3 - \frac{4}{3} \eta_1^3 \right) - 
 \frac{1}{\eta_1^2} \left( \xi_1^3 - \xi_1 \eta_1^2 \right)
\ln \left( \frac{\xi_1 + \eta_1}{\xi_1 - \eta_1} \right)\;,
\end{displaymath}
Herewith the notation $X_2 \left( \xi_1,\eta_1 \right)$ in the expressions for ${\cal D}_2$ and 
${\cal D}_3$ has the form 
\begin{equation}
X_2 \left( \xi_1,\eta_1 \right) = 4 \xi_1^2 - 4 \xi_1 - 4 \xi_1 \eta_1 + 2 \eta_1 + 3 \;. 
\end{equation} 
Using the formulas (24), the expressions for $X_2 \left( \xi_1,\eta_1 \right)$ and 
$X_2 \left( \xi_1,- \eta_1 \right)$ can be written in the form
\begin{equation}
X_2 \left( \xi_1,\eta_1 \right) = - (2 \xi_1 - 1) \cos \omega_{01} + 2 \;, \qquad
X_2 \left( \xi_1,- \eta_1 \right) = - (2 \xi_1 - 1) \cos \omega_{\pi 1} + 2 \;.
\end{equation} 
As in the preceding case of the $p$-wave component of the $t$-matrix, the first term in (40) 
is the Born approximation for the $d$-wave component,
\begin{equation}
t^{C,B}_2(k,k^{\prime};-b_1) = v^C_2(k,k^{\prime}) = 
\frac{\pi q_1 q_2}{k k^{\prime}} {\cal D}_1\;.
\end{equation} 

Adding up in (40) all four terms (41) and taking into consideration the relations 
(43), we obtain the formula for the $d$-wave Coulomb transition matrix at $E=-b_1$ :
\begin{displaymath}
t^C_2(k,k^{\prime};-b_1) = \frac{\pi q_1 q_2}{k k^{\prime}} \left\{  
\frac{1}{\eta_1} \left( 4 \xi_1^2 - 5 \xi_1 - \frac{8}{3} \eta_1^2 + \frac{3}{2} \right) 
\right.
\end{displaymath}
\begin{equation}
+ \frac{1}{\eta_1^2} \left[ \left( - \xi_1^3 + \frac{3}{2} \xi_1^2 + \xi_1 \eta_1^2 
- \frac{1}{2} \eta_1^2 \right) \ln \left( \frac{\xi_1 + \eta_1}{\xi_1 - \eta_1} \right) 
+ \frac{3}{16}  \left( 2 \xi_1 - 1 \right) \left( \omega_{\pi 1} - \omega_{01} \right)
 \left( 2\pi - \omega_{\pi 1} - \omega_{01} \right) \right.
\end{equation} 
\begin{displaymath}
\left. \left. + \frac{1}{8} \left( \pi - \omega_{\pi 1} \right) X_2 \left( \xi_1, \eta_1 \right) 
\sin \omega_{\pi 1} - \frac{1}{8} \left( \pi - \omega_{01} \right) X_2 \left( \xi_1, 
- \eta_1 \right) \sin \omega_{01} \right] \right\}\;,
\end{displaymath}
\\ [.2in]
\noindent {\sl 4.3. Partial $f$-wave Coulomb transition matrix 
 at the energy $E=-b_1$}\\

Likewise, integrating in the formula (33) with $l=3$, we determine the conributions 
from the individual terms in the braces to the $f$-wave Coulomb transition matrix:
\begin{equation}
t^C_3(k,k^{\prime};-b_1) = \frac{\pi q_1 q_2}{k k^{\prime}} \left\{ {\cal F}_1 
+ {\cal F}_2 + {\cal F}_3 + {\cal F}_4 \right\}\;,
\end{equation} 
where 
\begin{displaymath}
{\cal F}_1 = \frac{1}{\eta_1^2} \left( -5 \xi_1^2 + \frac{4}{3} \eta_1^2 \right) 
+ \frac{1}{2 \eta_1^3} \left( 5 \xi_1^3 - 3 \xi_1 \eta_1^2 \right) 
\ln \left( \frac{\xi_1 + \eta_1}{\xi_1 - \eta_1} \right) \;,       
\end{displaymath}
\begin{displaymath}
{\cal F}_2 = \frac{\pi}{16 \eta_1^3} \left[ 6 \left(5 \xi_1^2 - 5 \xi_1 - 
\eta_1^2 + \frac{25}{16} \right) \left( \omega_{\pi 1} - \omega_{01} \right) \right.
\end{displaymath}
\begin{displaymath}
\left. + X_3 \left( \xi_1,\eta_1 \right) \sin \omega_{\pi 1} - 
X_3 \left( \xi_1,- \eta_1 \right) 
\sin \omega_{01} \right] \;,
\end{displaymath}
\begin{equation}
{\cal F}_3 = \frac{1}{\eta_1^2} \left( \frac{5}{2}\xi_1^3 - \frac{15}{4} \xi_1^2 
+ \frac{95}{16} \xi_1 - \frac{13}{6} \xi_1 \eta_1^2 + \frac{13}{12} \eta_1^2 
- \frac{75}{32} \right) 
\end{equation}
\begin{displaymath}
+ \frac{1}{16 \eta_1^3} \left[ - 3 \left( 5 \xi_1^2 
- 5 \xi_1 - \eta_1^2 + \frac{25}{16} \right) 
\left( \omega_{\pi 1}^2 - \omega_{01}^2 \right) \right.
\end{displaymath} 
\begin{displaymath}
\left. - \omega_{\pi 1} 
X_3 \left( \xi_1,\eta_1 \right) \sin \omega_{\pi 1} +  \omega_{01} 
X_3 \left( \xi_1,- \eta_1 \right) \sin \omega_{01} \right] \;, 
\end{displaymath}
\begin{displaymath}
{\cal F}_4 = \frac{1}{\eta_1^2} \left( \frac{5}{2} \xi_1^3 - 
\frac{13}{6} \xi_1 \eta_1^2 \right) + \frac{1}{\eta_1^3} \left( - \frac{5}{4} \xi_1^4 
+ \frac{3}{2} \xi_1^2 \eta_1^2 - \frac{1}{4} \eta_1^4 \right)
\ln \left( \frac{\xi_1 + \eta_1}{\xi_1 - \eta_1} \right)\;,
\end{displaymath}
herewith the notation $X_2 \left( \xi_1,\eta_1 \right) $ in the expressions 
for ${\cal F}_2$ and ${\cal F}_3$ has the form 
\begin{equation}
X_3 \left( \xi_1,\eta_1 \right) = 10 \xi_1^3 - 15 \xi_1^2 + 
\frac{95}{4} \xi_1 - 10 \xi_1^2 \eta_1 + 10 \xi_1 \eta_1 -2 \xi_1 \eta_1^2 - 
\frac{25}{4} \eta_1 + \eta_1^2 + 2 \eta_1^3 - \frac{75}{8} \;. 
\end{equation} 
Using the formulas (24), the expressions for $X_3 \left( \xi_1,\eta_1 \right) $ and 
$X_3 \left( \xi_1,- \eta_1 \right) $ can be written in the form
\begin{displaymath}
X_3 \left( \xi_1,\eta_1 \right) = - \left( 5 \xi_1^2 - 5 \xi_1 - \eta_1^2 + 
\frac{75}{8} \right) \cos \omega_{01} + \frac{25}{2} \eta_1 \;,
\end{displaymath}
\begin{equation}
X_3 \left( \xi_1,- \eta_1 \right) = - \left( 5 \xi_1^2 - 5 \xi_1 - \eta_1^2 + 
\frac{75}{8} \right) \cos \omega_{\pi 1} - \frac{25}{2} \eta_1 \;.
\end{equation} 

The first addend in (46) is the Born approximation for the $f$-wave Coulomb $t$-matrix
\begin{equation}
t^{C,B}_3(k,k^{\prime};-b_1) = v^C_3(k,k^{\prime}) = 
\frac{\pi q_1 q_2}{k k^{\prime}} {\cal F}_1\;.
\end{equation} 

Adding up in (46) all four terms (47) and taking into consideration the relations 
(49), we derive the formula for the $f$-wave Coulomb transition matrix at the ground 
bound state energy $E=-b_1$ :
\begin{displaymath}
t^C_3(k,k^{\prime};-b_1) = \frac{\pi q_1 q_2}{k k^{\prime}} \left\{  
\frac{1}{\eta_1^2} \left( 5 \xi_1^3 - \frac{35}{4} \xi_1^2 + \frac{95}{16} \xi_1 - 
\frac{13}{3} \xi_1 \eta_1^2 + \frac{29}{12} \eta_1^2 - \frac{75}{32} \right) \right.
\end{displaymath}
\begin{displaymath}
+ \frac{1}{\eta_1^3} \left[ \left(- \frac{5}{4} \xi_1^4 + \frac{5}{2} \xi_1^3 
+ \frac{3}{2} \xi_1^2 \eta_1^2 - \frac{3}{2} \xi_1 \eta_1^2 - \frac{1}{4} \eta_1^4 \right)
\ln \left( \frac{\xi_1 + \eta_1}{\xi_1 - \eta_1} \right) \right.
\end{displaymath} 
\begin{equation}
+ \frac{3}{16} \left( 5\xi_1^2 - 5 \xi_1 - \eta_1^2 + \frac{25}{16} \right)
\left( \omega_{\pi 1} - \omega_{01} \right) \left( 2 \pi - \omega_{\pi 1} 
- \omega_{01} \right) 
\end{equation} 
\begin{displaymath}
\left. \left. + \frac{1}{16} X_3 \left( \xi_1, \eta_1 \right) \left( \pi - 
\omega_{\pi 1} \right) \sin \omega_{\pi 1} - \frac{1}{16} 
X_3 \left( \xi_1, - \eta_1 \right) \left( \pi - \omega_{01} \right) 
\sin \omega_{01} \right] \right\}\;,
\end{displaymath}
\\ [.2in]
\noindent {\bf 5. Concluding remarks}\\

In this work we have developed the method of the analytical solving the integral 
Lippmann-Schwinger equation for the partial wave off-shell two-body Coulomb 
transition matrix at the energy of the ground bound state, $t^C_l(k,k^{\prime};-b_1)$ 
( $l=1,2,3,...$), using the Fock's dynamical symmetry of the system with the Coulomb 
interaction described by the four-dimensional rotation group SO(4). 

Specifically, a knowledge of the partial-wave Coulomb transition matrix at the  
bound state energy is necessary to determine the electric multipole polarizabilities 
of the hydrogen-like atoms. 

Restricting our attension to the hydrogen-like atom with an infinite-mass nucleus we 
write the expression for the electric $2^\lambda$-pole polarizabilities of the two-body 
bound complex $\alpha_{E\lambda}$ ($\lambda=1,2,3,...$) 
in terms of the wave function, its corresponding derivatives and the partial wave 
transition matrix $t_{\lambda}$ at the energy of the bound state $E=-b_1$ (Ref. 16): 
\begin{displaymath}
\alpha_{E\lambda} = \frac{2}{(2\lambda+1)\pi^2} \frac{m_1{q_1}^2}{\hbar^2}
\left\{ \int_{0}^{\infty} dk k^2 \frac{\mid \varphi_{\lambda}(k)
\mid ^2}{k^2+\kappa_1^2} \right. \\ [-2mm] \nonumber
\end{displaymath} 
\begin{equation}
  - \frac{m_1}{\pi^2 \hbar^2} \int_{0}^{\infty}dk k^2 
\int_{0}^{\infty} dk^{\prime}k^{{\prime}2} 
\left. \frac{\varphi^{*}_{\lambda}(k) t_{\lambda}(k,k^{\prime};-b_1)
\varphi_{\lambda}(k^{\prime})}
{(k^2+\kappa_1^2)(k^{{\prime}2}+\kappa_1^2)} \right\}\;, 
\end{equation}
where $m_1$ and $q_1$ are the mass and the charge of the electron,
\begin{equation}
\varphi_{\lambda}(k) = (-1)^{\lambda} k^{\lambda} \left[ \left( \frac{1}{k} \frac{d}{dk} 
\right) ^{\lambda} \psi(k)\right]\;,
\end{equation}
$\psi(k)$ is the wave function of the $S$-wave ground bound state of the atom in the 
momentum space. 

Introducing into consideration a so-called {\it off-shell-energy scattering function }
\begin{equation}
\phi_{\lambda}(k) = -\frac{m_1}{\pi^2 \hbar^2} \int_{0}^{\infty} dk^{\prime} {k^{\prime}}^2 
t^C_{\lambda}(k,k^{\prime};-b_1) \frac{1}{{{k^{\prime}}^2}+{\kappa_1}^2} 
\varphi_\lambda(k^{\prime}) 
\end{equation}
the integral equation for which follows from the Lippmann-Schwinger equation for 
the partial transition matrix at $E=-b_1$ (19),
\begin{equation}
\phi_{\lambda}(k) = f_{\lambda}(k) -\frac{m_1}{\pi^2 \hbar^2} \int_{0}^{\infty} 
dk^{\prime} {k^{\prime}}^2 v^C_{\lambda}(k,k^{\prime})\frac{1}{{{k^{\prime}}^2}+{\kappa_1}^2} 
\phi_{\lambda}(k^{\prime}) \;,
\end{equation}
where the free term is determined by the expression
\begin{equation}
f_{\lambda}(k) = -\frac{m_1}{\pi^2 \hbar^2} \int_{0}^{\infty} dk^{\prime} {k^{\prime}}^2 
v^C_{\lambda}(k,k^{\prime})\frac{1}{{{k^{\prime}}^2}+{\kappa_1}^2}  
\varphi_{\lambda}(k^{\prime}) \;, 
\end{equation}
we write the formula (52) as
\begin{equation}
\alpha_{E\lambda} = \frac{2}{(2\lambda+1)\pi^2} \frac{m_1 {q_1}^2}{\hbar^2}
\int_{0}^{\infty} dk k^2 \frac{\varphi_{\lambda}^{*}(k) \left[ \varphi_{\lambda}(k) 
+ \phi_{\lambda}(k) \right]}{k^2 + \kappa_1^2} \;.
\end{equation}

In the preceding paper [17] the integral equation (55) for the 
hydrogen-like atom has been analytically solved. It has also been 
shown that the formula (57) reproduces the analytical result for the multipole 
electric polarizability derived by Dalgarno and Lewis in Ref. [18].

The method of the derivation of the analytical expression for 
the partial wave Coulomb transition matrices developed in this 
work can be generalized for determination of the transition 
matrix at the excited energies, $t^C_l(k,k^{\prime};-b_n)$ with $n>1$, 
the knowledge of which, in particular, makes it possible to obtain information 
on the electric polarizabilities of the hydrogen-like atom in the 
excited states. 

In nuclear physics, the direct approach using the off-shell 
transition matrix has been applied in [19] to calculate 
the electric multipole polarizabilities of the deuteron.
\\ [.2in]

\noindent {\bf Acknowledgment}\\

The present work was partially supported by the National Academy of Sciences 
of Ukraine (project No. 0112U000054) and by the Program of Fundamental Research 
of the Department of Physics and Astronomy of NASU (project No. 0112000056). 
\\ [.2in] 
\clearpage
\noindent {\footnotesize {\bf References} 
\vspace*{.1in}
\begin{itemize} 
\setlength{\baselineskip}{.1in} 
\item[{\tt [1]}] L.D.Faddeev, Sov. Phys. JETP 12(1961)1014-1019. 
\item[{\tt [2]}] L.D.Faddeev, Mathematical Aspects of the Three-Body Problem in the Quantum
                 Scattering Theory. Isr. Program Sci. Transl., Jerusalem, 1965. 
\item[{\tt [3]}] S.Okubo, D.Feldman, Phys. Rev. 117(1960)292-306.
\item[{\tt [4]}] E.H.Wichmann, C.H.Woo, J. Math. Phys. 2(1961)178-181.
\item[{\tt [5]}] V.F.Bratsev, E.D.Trifonov, Vest. Leningrad. Gos. Univ. 16(1962)36-39. 
\item[{\tt [6]}] L.Hostler, J. Math. Phys. 5(1964)591-611.
\item[{\tt [7]}] L.Hostler, J. Math. Phys. 5(1964)1235-1240.
\item[{\tt [8]}] J.Schwinger, J. Math. Phys. 5(1964)1606-1608.
\item[{\tt [9]}] A.M.Perelomov, V.S.Popov, Sov. Phys. JETP 23(1966)118-134.  
\item[{\tt [10]}] S.A.Shadchin, V.F.Kharchenko, J. Phys. B 16(1983)1319-1322. 
\item[{\tt [11]}] S.A.Storozhenko,S.A.Shadchin, Teor. Mat. Fiz. 76(1988)339-349. 
\item[{\tt [12]}] H.van Haeringen, J. Math. Phys. 25(1984)3001-3032.
\item[{\tt [13]}] J.C.Y.Chen, A.C.Chen, Advances in Atomic and Molecular Physics, Vol.8, 
                  Ed. D.B.Bates and I.Estermann, Academic Press, N.Y.-London, 1972, p.p.71-129.
\item[{\tt [14]}] V.A.Fock, Z. Phys. 98(1935)145-154. 
\item[{\tt [15]}] I.S.Gradstein, I.M.Ryzhik, Tables of Integrals, Sums,Series and Products, 
                  Nauka, Moscow, 1971.
\item[{\tt [16]}] V.F.Kharchenko, J. Mod. Phys. 4(2013)99-107.
\item[{\tt [17]}] V.F.Kharchenko, Ann.Phys.NY 355(2015)153-169.
\item[{\tt [18]}] A. Dalgarno, J.T.Lewis, Proc. Roy. Soc. A233(1955)70-74.
\item[{\tt [19]}] V.F.Kharchenko, Int. J. Mod. Phys. E22(2013)1350031,1-15.

\end{itemize}} 

\end{document}